\documentclass[times,twocolumn]{aastex62}

\newcommand{\unitvec}[2]{\hat{\mathbf{#1}}_{#2}}
\newcommand{\texthide}[1]{}
\usepackage{natbib}
\usepackage{xspace}
\newcommand{\rev}[1]{#1}

\defcitealias{Marzari2002}{M\&W02}
\defcitealias{Simbulan2017}{S17}
\usepackage{subfigure}
\usepackage{savesym}
\savesymbol{tablenum}
\usepackage{siunitx}
\usepackage{amsmath}
\usepackage{datetime}

\usepackage[encapsulated]{CJK}
\usepackage{ucs}
\usepackage[utf8x]{inputenc}
\newcommand{\cntext}[1]{\begin{CJK*}{UTF8}{gbsn}#1\end{CJK*}}
\newcommand{\jptext}[1]{\begin{CJK*}{UTF8}{min}#1\end{CJK*}}

\restoresymbol{SIX}{tablenum}

\usepackage{ulem}
\newcommand{\rmv}[1]{}

\received{}
\revised{}
\accepted{}
\submitjournal{ApJ}
\shorttitle{Three-planet systems from the HL Tau disk}
\shortauthors{Wang et al.}

\begin{document}

\title{Architecture of three-planet systems predicted from the observed protoplanetary disk of HL Tau} \correspondingauthor{Shijie Wang} 
\email{wang@utap.phys.s.u-tokyo.ac.jp} 
\author[0000-0002-5635-2449]{Shijie Wang (\cntext{汪士杰})} 
\affil{Department of Physics, The University of Tokyo, Tokyo 113-0033, Japan} 

\author[0000-0001-7235-2417]{Kazuhiro D. Kanagawa (\jptext{金川和弘})} 
\affiliation{Research Center for the Early Universe, School of Science, Tokyo 113-0033, Japan} 

\author[0000-0003-0288-6901]{Toshinori Hayashi (\jptext{林利憲})} 
\affiliation{Department of Physics, The University of Tokyo, Tokyo 113-0033, Japan} 

\author[0000-0002-4858-7598]{Yasushi Suto (\jptext{須藤靖})}
\affiliation{Department of Physics, The University of Tokyo, Tokyo 113-0033, Japan} \affiliation{Research Center for the Early Universe, School of Science, Tokyo 113-0033, Japan}

\begin{abstract}
A number of protoplanetary disks observed with ALMA \rev{potentially} provide direct examples of initial conditions for planetary systems. In particular, the HL Tau disk has been intensively studied, and its rings/gaps are conventionally interpreted to be a result of unseen massive planets embedded in the gaps. Based on this interpretation, we carried out N-body simulations to investigate orbital evolution of planets within the protoplanetary disk and after the disk dispersal. Before the disk dispersal, our N-body simulations include both migration and mass-growth of the planet coupled with evolution of the disk. By varying the disk parameters, we produce a variety of widely-separated planetary systems consisting of three super-Jupiters at the end of disk dispersal. We found the outer planet is more massive than the inner one, and the migration of the innermost planet is inefficient due to the accretion of outer planet(s). We also showed how the final configuration and the final planetary mass depend on disk parameters. The migration is found to be convergent and no planet-pair has a period ratio less than \num{2}. After the disk dispersal, we switch to pure gravitational N-body simulations and integrate the orbits up to \SI{10}{Gyr}. Most simulated systems remain stable for at least \SI{10}{Gyr}. We discuss implications of our result in terms of the observed widely-separated planetary systems HR 8799 and PDS 70.
\end{abstract}


\section{Introduction}

Recent analyses \citep[e.g.,][]{Winn2014} on Kepler systems have revealed that the population of the exoplanets exhibited a great diversity. The existence of close-in gas giants \citep[e.g.][]{Mayor1995}, planets in extremely eccentric orbit \citep[e.g.][]{Jones2006} and planets with large spin-orbit misalignment \citep[e.g.][]{Narita2009,Winn2009} indicates that exoplanets have a broad distribution of planetary mass, orbital eccentricity and inclination. Such a diversity of the observed systems has not been expected from a conventional model for the Solar system \citep{Hayashi1985}, which has triggered a significant amount of theoretical and observation interests to identify the physical origin in the context of the initial condition and later evolution of exoplanets.

The observed diversity of the exoplanets may be explained, \rev{in part,} by chaotic dynamical evolution and orbital instability of the planets \citep[e.g.,][]{Rasio1996,Wu2003,Nagasawa2008,Naoz2011}. However, the stability of planetary systems depends on their configurations in terms of masses and orbital separations, which are determined by the final outcome of the orbital migration and mass growth of the planet within the protoplanetary disk. From this point of view, the stability of the planetary system is closely related to the evolution of the planet within the protoplanetary disk.

Planetary migration was originally proposed before the discovery of Hot Jupiters \citep{Lin1979,Goldreich1980}, and is regarded as a basic outcome of the gravitational interaction between a planet and a protoplanetary disk (PPD). More importantly, a planet accretes mass in the course of migration, and eventually opens a deep density gap in the PPD if it becomes sufficiently massive \citep{Lin1993,Paardekooper2004,Crida2006,Kanagawa2015}. Such a gap in the PPD, therefore, can offer potential evidence for proto-planet(s).

A significant amount of observational work has been devoted to the detection of PPDs and the analysis of their properties, including PPDs in Orion nebula \citep{Odell1993}, PPD surrounding HD 142527 \citep{Fukagawa2006} and PPDs in Orionis Cluster \citep{Hernandez2007}. Indeed, \cite{Partnership2015} identified the ring/gap substructure in the HL Tau dust disk clearly for the first time, with Atacama Large Millimeter Array (ALMA)\footnote{To avoid confusion, throughout the paper we use the terms ``ring'' and ``gap'' to denote bright and dark annulus regions in a disk, respectively.}.  As of October 2019, its follow-up project DSHARP \citep[e.g.][]{Andrews2018,Huang2018,Kurtovic2018} has released continuum data for 20 PPDs with similar ring/gap substructures. \cite{Keppler2018} and \cite{Haffert2019} also announced the discovery of two accreting planets inside the PDS 70 transiting disk. We can adopt masses and orbital radii of planets estimated from the observed gap/ring structures in the protoplanetary disk as initial conditions of our simulations for planetary evolution.

We also note that the dust gaps may be explained in alternative interpretations including \rev{secular gravitational instability} \citep{Takahashi2016}, snowlines \citep{Zhang2015}, sintering \citep{Okuzumi2016} and non-ideal MHD effect \citep[e.g.][]{Flock2015,Hu2019}. The planet interpretation, however, is widely believed to be a promising explanation. Thus we adopt this interpretation, and examine the future outcome of the multi-planetary systems predicted from the HL Tau disk in what follows.

Since the location and width of a gap along with the disk properties can be translated into the mass of a planet there \citep[e.g.][]{Kanagawa2016}, the initial conditions of the corresponding planets in the HL Tau disk \rev{can be reasonably constrained.} The pioneering work from \cite{Simbulan2017} (hereafter \citetalias{Simbulan2017}) is based on this idea. They assigned four or five planets at the locations of the gaps, and performed a series of numerical simulations to predict the fate of the observed HL Tau system. \citetalias{Simbulan2017} found that more than half of the planets were ejected from the system, collided with another planet, and/or migrated within the Roche radius of the host star. Their conclusion is interesting and may explain the origin of the observed diversity of exoplanets including eccentric cold Jupiters, hot Jupiters, and free-floating planets, at least in a qualitative fashion.

There are a few points in \citetalias{Simbulan2017} that need to be carefully checked, especially the assumption that inward migration induced by the disk-planet interaction is quite inefficient. Since the age of the HL Tau is only around \SI{1}{Myr}, the migration and mass accretion of the planets during the rest of the disk lifetime are expected to play an important role in the later stage orbital evolution. This motivates us to improve their work by performing systematic simulations in the present paper to examine the fate and long-term stability of planetary systems predicted from the current HL Tau disk.

The rest of the paper is organized as follows. Observed properties of the HL Tau system are summarized in section \ref{sec:hltau}. Section \ref{sec:methods} describes our methodology, including equations of motion, models of migration and accretion, model of disk profile evolution, and initial conditions. We present results of our simulations in section \ref{sec:rslt}, and discuss implications of our results in section \ref{sec:discussion} with particular attention to the orbital stability of the multi-planets. Section \ref{sec:summary} is the summary of this paper.

\section{The HL Tau system \label{sec:hltau}}

HL Tau is a young star hosting a PPD in the Taurus star forming region. It is a well-studied system for interferometric observation even prior to ALMA because of its high brightness at millimetre wavelength. The HL Tau disk is one of the most massive disks that have been observed to date. The disk mass has been estimated to be between $ \SI{0.03}{M_\odot} $ and $ \SI{0.14}{M_\odot}$ \citep{Robitaille2007,Guilloteau2011,Kwon2011}, and the recent result from \citet{Kwon2015} gives the mass of $\SI{0.105}{M_\odot}$. \rev{It should be noted that the disk mass is indirectly inferred from the opacity estimated from the thermal emission of the dust grains. The total mass of the dust grains, therefore, depends on the opacity of the dust grain and its size distribution as well, which are not well constrained. Moreover, we need to specify a gas-to-dust ratio to translate the dust mass to the gas mass, which is also poorly understood and often assumed to be a constant. Due to the above uncertainties, the estimate of the disk mass may vary in an order of magnitude.}

The HL Tau system attracted particular attention because it is the first PPD whose substructure is well resolved by ALMA high angular resolution observation \citep{Partnership2015}. ALMA successfully resolved the dust concentric ring/gap substructure and identified seven bright rings and dark gaps. Follow-up modelling on the continuum data performed by \cite{Pinte2015} shows the deepest three major gaps are located at \SI{13.2}{au}, \SI{32.3}{au} and \SI{73.7}{au}, in which the dust density is depleted to at least \rev{a factor of \num{10}.} The other four gaps are relatively shallow.

Since the inclination angle of the HL tau disk is around $\SI{47}{\degree} $, ALMA images also resolve the vertical structure of the disk in high resolution. Based on the fact that the gaps and rings are sharp at all azimuthal angles, \cite{Pinte2015} concluded that the HL Tau disk is geometrically-thin and that the majority of the dust is settled near the disk mid-plane. Such a shape suggests a weak turbulence level inside the disk, and thus, a low viscosity due to ineffective angular momentum transfer. \cite{Pinte2015} shows that $\alpha = \num{3e-4}$ for the $\alpha$-parameter on the viscosity \citep{Shakura1973} well reproduces the observed image of the HL Tau disk.

In this paper, we adopt the most conventional interpretation that the gaps are caused by planets, and consider the evolution of planets implied from the HL Tau disk. \rev{However, even the planet interpretation is adopted, many uncertainties are still involved while mapping the number of gaps to the number of planets hosting by HL Tau. For example, \cite{Dong2018} have shown that a single planet can produce multiple gaps in the case of low viscosity, showing the number of planets can be less than the number of gaps. By considering the early debris disk dynamics, \cite{Boley2017} also pointed out that three giant planets can produce both major and minor gaps of HL Tau via dynamical interactions between planets and planetesimals. For simplicity, we consider the case of three proto-planets in our simulations. The three-planet interpretation is also supported by the results of hydrodynamic simulations \citep[e.g.,][]{Dipierro2015,Picogna2015,Jin2016}, in which the major gap structures can be reproduced by three planets.}We present the properties and initial configuration of the planets in our simulations in section \ref{sec:inicons}.

\section{Methods \label{sec:methods}}

In this section, we present our methods to evolve the HL Tau planetary
system from the disk stage to \SI{10}{Gyr} after the disk dispersal.
We also describe the equation of motion for planets inside a disk as well as our
accretion and migration model of planets coupled with a surrounding
disk.

\subsection{Equation of motion of planets \label{sec:eqn}}

For a planet within a disk, in addition to the gravitational forces
exerted by the central star and other planets, the planet will
experience forces due to planet-disk interactions. Consider the forces
acting on the \textit{i}-th planet, the equation of motion can be
written as
\begin{equation}\label{EOM_pri}
  \ddot{\boldsymbol{r}}_{i} = \boldsymbol{f}_{\text{grav},i}
  + \boldsymbol{f}_{\text{a},i} + \boldsymbol{f}_{\text{e},i},
\end{equation}
where $ \boldsymbol{r}_i $ is the position vector of the \textit{i}-th
planet. The $ \boldsymbol{f} $ notations on the right hand side of the
equation are respective forces \textit{per unit mass} exerting on the
\textit{i}-th planet. We assume the planets to be co-planar, and thus $
\boldsymbol{r}_i $ can be replaced by the position vector $
\boldsymbol{R}_i(R_i,\phi_{i})$ in cylindrical coordinate system. We
use $\boldsymbol{R} $ to denote the cylindrical position vector unless
stated otherwise.

The first term, $\boldsymbol{f}_{\text{grav}}$, of the right-hand-side
of equation (\ref{EOM_pri}) denotes gravitational force exerted by the
central star and other planets, and is given by
\citep[e.g.][]{murray2000solar}:\newpage
\begin{align}\label{6.1fgrav}
&\boldsymbol{f}_{\text{grav},i}
 = -\mathcal{G}(M_*+M_i)\frac{\boldsymbol{R}_i}{R_i^3}  \cr
 &~ +\sum_{k}^{k\neq i}\frac{\mathcal{G}M_k}{\parallel \boldsymbol{R}_k
   - \boldsymbol{R}_i \parallel^3}(\boldsymbol{R}_k - \boldsymbol{R}_i)
 -\sum_{k}^{k\neq i} \frac{\mathcal{G}M_k}{R_k^3}\boldsymbol{R}_k,
\end{align}
where $M_*$ is the mass of the central star, $M_i$ is the mass of the
$i$-th planet, $R = \lvert\boldsymbol{R}\rvert$, and $\mathcal{G}$ is
the gravitational constant.

The second term, $ \boldsymbol{f}_{a,i}$, denotes the force driving
the migration of planet. A planet with index $i$ embedded in the disk
experiences an effective torque $\Gamma_i$, and migrates inwards.  To
describe the motion, it is convenient to introduce an e-folding inward
migration time-scale $\tau_{a,i}$:
\begin{equation}
  \label{6.1taua_intuitive}
 \tau_{a,i}(t) \equiv -\frac{a_i}{{\rm d}a_i/{\rm d}t},
\end{equation}
where $a_i$ is the semi-major axis of the $i$-th planet. In the case
of a co-planar and near circular orbit, $|\boldsymbol{R}_{i} | \approx
a_i$, and equation (\ref{6.1taua_intuitive}) reduces to
\begin{equation}
\tau_{a,i} = \frac{L_{i}}{2\Gamma_i}
\end{equation}
in terms of the angular momentum $L_{i}$ and the torque $ \Gamma_i $ acting on the planet from the disk.
Thus, the effective migration force, $ \boldsymbol{f}_{a,i} $, is
simply given by
\begin{equation}
  \boldsymbol{f}_{\text{a},i} = \ddot{\boldsymbol{R}}_{i}
  = -\frac{\dot{\boldsymbol{R}}_{i}}{2\tau_{a,i}}.
\end{equation}
The disk tends to circularise a planetary orbit.
Similarly  as equation (\ref{6.1taua_intuitive}),
we define the eccentricity damping time-scale, $\tau_{e,i}$.
Following \citet{Hoi2002,Kley2004}, we further assume the
following expression:
\begin{equation}\label{6.1eccdamping}
\tau_{e,i} = C\tau_{a,i}\left(\frac{h_i}{R_i}\right)^2,
\end{equation}
where $C$ is a constant, and $h_i/R_i$ is the aspect ratio of the disk
at the location of the $i$-th planet.
Then the eccentricity damping force is computed as
\begin{equation}
\label{6.1wiley_edamp}
  \boldsymbol{f}_{\text{e},i}
  = -\frac{2}{3\tau_{e,i}}\left(\frac{\dot{\boldsymbol{R}}_{i}}{1-e_i^2}
  - \unitvec{\boldsymbol\ell}{i}\times
  \unitvec{\boldsymbol R}{i}
  \sqrt{\frac{\mathcal{G}(M_*+M_i)}{a_i(1-e_i^2)}}\right), 
\end{equation}
where $\unitvec{\boldsymbol\ell}{i} $ is the unit vector of the
specific angular momentum of the $ i $-th planet. \rev{(See Appendix A of \cite{Hoi2002}).}

\begin{figure*}
\centering
\includegraphics[width=0.8\linewidth]{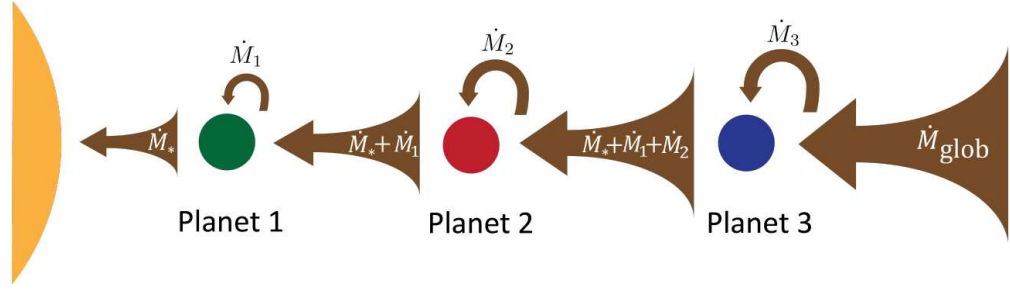}
\caption{Schematic view of the mass flow from outside of the disk to
  the inside. Blue circles and green arrows are the planets and mass
  flow. Due to the accretion of the planet, mass flow is discontinuous
  at the position of each planet.}
	\label{fig:schematicflow}
\end{figure*}

\subsection{Migration model}

In section \ref{sec:eqn} we have defined the migration time-scale $
\tau_{a,i}$ to capture the physics of the migration process.  In
practice, we adopt an empirical model $\tau_{a,i}$ by
\citet{Kanagawa2018}.  Since this is one of the most important
processes that we implement in the current simulation, we summarise
the explicit formulae in this subsection.

\cite{Kanagawa2018} performed two-dimensional hydrodynamical simulations, and
investigated the planetary migration by varying disk parameters.  They
found that the migration slows down as the gap becomes deeper. The
depth of the gap, defined as the ratio between the minimum bottom
density of the gap $\Sigma_{{\rm min},i}$ and the surface density of the
unperturbed vicinity $ \Sigma_g(R_i) $, can be characterised by a
dimensionless factor $K_i $ as
\begin{align}
\frac{\Sigma_{min,i}}{\Sigma_g(R_i)} = \frac{1}{1+0.04K_i},
\label{6.1siggap}
\end{align}
where
\begin{align}
\label{eq:def-K}
K_i = \left(\frac{M_i}{M_*}\right)^2
\left(\frac{h_i}{R_i}\right)^{-5}\alpha^{-1} .
\end{align}
A small value of $K_i$ corresponds to the case of a small planet mass
($M_i$), in which the planetary migration is in Type I regime.
As the planet mass grows due to accretion from disk, the gap depth and
therefore $K_i$ gradually increase, and the migration becomes slower
(Type II regime).

Adopting this gap parametrisation, \citet{Kanagawa2018} found the
following empirical expression of $\tau_{a,i}$ that incorporates the
slower migration for the deeper gap:
\begin{equation}
  \label{5.2taui_tsc}
  \tau_{a,i} = \dfrac{1+0.04K_i}{\gamma_{L,i}
    + \gamma_{C,i}\exp(-K_i/K_{t})}\tau_{0,i}(R_i).
\end{equation}
In equation (\ref{5.2taui_tsc}), $\gamma_{C,i}=\Gamma_{C,i}/\Gamma_{0,i}$ and $\gamma_{L,i}=\Gamma_{L,i}/\Gamma_{0,i}$, where $ \Gamma_{C,i}$, $\Gamma_{L,i}$, and $\Gamma_{0,i}$ denote the co-rotation, Lindblad, and characteristic torque for the $i$-th planet, respectively. The characteristic torque $ \Gamma_{0,i} $ is defined as
\begin{align}\label{eqn:cgamma11}
  \Gamma_{0,i}(R_i) = \left(\frac{M_i}{M_*}\right)^2
  \left(\frac{h_i}{R_i}\right)^{-2}\Sigma_g(R_i) R_i^4\Omega^2_{K,i}.
\end{align}
Here $K_{t}$ is the co-rotational cut-off value of $K_i$, $\Omega_{K,i} $ is the Keplerian angular velocity of the $i$-th planet, and $\tau_{0,i}$
is the characteristic time-scale defined as $ \tau_{0,i} =
L_i/(2\Gamma_{0,i})$.
For a deep gap ($ K_i \gg K_t$), equation (\ref{5.2taui_tsc}) reduces
to
\begin{equation}\label{taua_approx}
  \tau_{a,i} \approx
  \frac{1+0.04K_i}{\gamma_{L,i}}\tau_{0,i}(R_i),
\end{equation}
which reflects the fact that the migration becomes slower for a deeper
gap.

We further adopt the following explicit expressions for the
dimensionless torques, which are derived by \citet{Paardekooper2011}
assuming that the disk is locally isothermal:
\begin{align}
\gamma_{C,i}&= 1.1(1.5-s)b_i+2.2\beta b_i^{0.71}-1.4\beta b_i^{1.26},\\
\gamma_{L,i}&= -(2.5-0.1s+1.7\beta)b_i^{0.71}\label{eqn:gammal14},
\end{align}
where $ s = -\ln \Sigma_g/\ln R $ and $\beta = -\ln T/\ln R$ are the power-law indices of the surface density and temperature profiles. The dimensionless factor $ b_i $ was originally introduced by \cite{Paardekooper2011}. The value of $ b_i$ is fixed to be $ 2/3 $ in our simulation, following \cite{Kanagawa2018}.

\subsection{Mass accretion model}

\cite{Tanigawa2016} studied the mass growth of a giant planet and obtained a semi-empirical model of the gas accretion rate onto the planet, which is in reasonable agreement with the results of hydrodynamic simulations. We adopt their model of planetary accretion and hereby summarise key expressions relevant to our simulation.

The model of \cite{Tanigawa2016} states that the accretion rate of the $i$-th planet, $\dot{M}_i$, can be expressed in terms of the accretion \textit{area} per unit time $D_i$ and the disk surface density of the accretion channel $\Sigma_{{\rm acc}, i}$ at its location $R_i$. Since the gas accretion only takes place at the close vicinity of the planet (gas can accrete at around two Hill radii from the planet), we can replace $\Sigma_{\rm acc,i}$ to $\Sigma_{\rm min,i}$ given by Equation~(\ref{6.1siggap}). $\dot{M}_i$ is then written as\footnote{Instead of Equation~(\ref{6.1siggap}), \cite{Tanigawa2016} adopt a slightly different formula $\Sigma_{\rm min,i} = 1/(1+0.034K_i)$. This difference does not change our main results.}
\begin{gather}
\label{6.1accmodel}
\dot{M}_i = D_i\Sigma_{{\rm min},i}, \\
D_i = 0.29\left(\frac{h_i}{R_i}\right)^{-2}
\left(\frac{M_i}{M_*}\right)^{4/3}R_i^2\Omega_{K,i},
\end{gather}
We compute $\dot{M}_i$ iteratively for a given $\dot{M}_{\rm glob}$ and other planetary configuration.
\subsection{Model of the disk profile hosting multiple planets}
\label{method_disk_structure}

The planetary migration and accretion require a disk profile model. For a static and axisymmetric disk surrounding a star, the surface density profile is given by
\begin{equation}
  \label{unp_disk_prof}
  \Sigma_{g}(R) =
  \frac{\dot{M}_*}{3\pi\nu(R)}\left(1-\sqrt{\frac{R_*}{R}}\right),
\end{equation}
where $\dot{M}_*$ is the stellar accretion rate, and $\nu$ is the
kinematic viscosity \citep{Pringle1981}.

Equation (\ref{unp_disk_prof}) assumes that the disk mass accretes on
the central star only and neglects the accretion on planets.
\cite{Tanigawa2016} incorporates the mass accretion onto a single
planet by approximating the planet as a sink for mass and angular
momentum within an inward disk mass flow. We generalise their model
for a multi-planetary system, as schematically illustrated in Figure
\ref{fig:schematicflow}.

To be more specific, we consider a quasi-steady disk with the $i$-th
planet located at $R_i$ with its accretion rate $\dot{M}_i$ ($i= 1,2,
\cdots, N$). We define a global accretion rate as the sum of all the
accretion rates:
\begin{equation}\label{6.1mglob}
\dot{M}_{\rm glob} = \dot{M}_*+\sum\limits_{i=1}^{N}\dot{M}_i.
\end{equation}
Moreover we assume that the initial value of $ \dot{M}_{\rm glob} $
only depends on global parameters of the disk, which are unaffected by
the migration and accretion of the planets (see later section
\ref{sec:inicons}). Then the surface density profile
$\Sigma_{\rm g}(R)$ between the $n$-th and $(n+1)$-th planets
($R_n \leq R<R_{n+1}$) is modelled as
\begin{align}
  \Sigma_g(R) = \frac{\dot{M}_*}{3\pi\nu}\left(1-\sqrt{\frac{R_*}{R}}\right)
  + \sum\limits_{i=1}^{n}\frac{\dot{M}_i}{3\pi\nu}
  \left(1-\sqrt{\frac{R_i}{R}}\right) 
\label{multip_sigmaunp}
\end{align}
(see Appendix \ref{sec:appendix1}). The above formula shows that the surface density profile couples with the migration and accretion of the planet. Particularly, when $R$ is large, the profile reduces to the form of equation (\ref{unp_disk_prof}), \textit{i.e.}, $\Sigma_g(R) \simeq \dot{M}_{\rm glob}/(3\pi\nu)$.

Since the location that we consider is far away from the star ($R \gg R_*$), the bracket of the first term on the right-hand-side of equation~(\ref{multip_sigmaunp}) is unity in practice. If the position and mass of each planet are specified, we can first express the accretion rate of each planet in terms of $\dot{M}_{\rm glob}$ by evaluating equation~(\ref{multip_sigmaunp}) at each $ R_i $ and then substituting to equation~(\ref{6.1accmodel}). Once $\dot{M}_{\rm glob}$ is given, we can then compute the explicit mass accretion rate of each planet as well as surface density at an arbitrary distance $ R $.

We take account of the gas removal from the disk, simply adopting an exponentially decaying model:
\begin{equation}\label{6.1diskdecay}
\dot{M}_{\rm glob}(t) = \dot{M}_{\rm glob}^{\rm ini}e^{-t/\tau_{\rm disk}},
\end{equation}
where $\dot{M}_{\rm glob}^{\rm ini}$ is the initial value of the
global mass acrretion rate, and $\tau_{disk}$ is the e-folding decay time of
the disk mass.

Due to the large uncertainty of the disk lifetime, we treat $\tau_{\rm disk}$ as a free parameter in the simulation and consider $\tau_{\rm disk}=1$, 2 and 3 Myr. Note that for simplicity, the surface density profile given by equation~(\ref{multip_sigmaunp}) does not include the shape of the gap induced by the planet. Equation~(\ref{multip_sigmaunp}) gives $\Sigma_{\rm g} (R_i)$ in equation~(\ref{6.1siggap}).

\begin{figure*}
	\centering
	\includegraphics[width=0.8\linewidth]{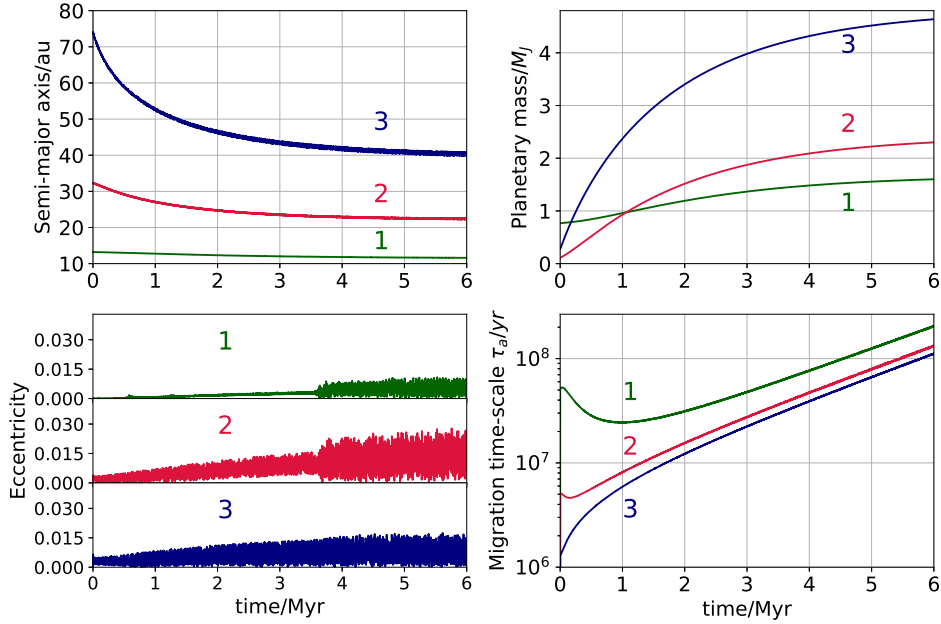}
	\caption{\rev{The evolution of semi-major axis(top left), planetary mass(top right), eccentricity(bottom left) and migration time-scale(bottom right) in the fiducial case ($ f = 0.25 $, $ \tau_{disk} = \SI{2}{Myr} $, $ \alpha = \num{3e-4} $). The numbers are the planet indices. }}
	\label{fig:amet_evo}
\end{figure*}

\subsection{Numerical method and initial conditions}
\label{sec:inicons}
\begin{table}
	\centering
	\caption{Initial conditions of HL Tau disk}
	\resizebox{\linewidth}{!}{%
		\begin{tabular}{llc}
			\hline
			\hline
			{\textbf{Notation}} & {\textbf{Meaning}} & {\textbf{Value}} \\
			\hline
			{\textbf{Free Parameters}}&&\\
			$ f $& {Flaring index} &[$0.15$,$0.20$,$\underline{\boldsymbol{0.25}}$,$0.30$,$0.35$]\\
			$ \tau_{\rm disk} $&{Disk lifetime}& $ [1,\underline{\boldsymbol{2}},3]\si{Myr} $ \\
			$ \alpha $ & {Viscosity parameter} &$ [2,\underline{\boldsymbol{3}},4,5,6]\times \num{e-4}$ \\   
			\hline
			{\textbf{Fixed Parameters}}&&\\
			$ M_{\rm disk}^{\rm ini} $& {Initial disk mass}&$ \SI{0.105}{M_\odot} $\\
			$ M_* $&{Stellar mass} & $ \SI{1.0}{M_{\odot}}$\\
			$ \left(h/R\right)_{\SI{1}{au}} $&{Aspect ratio at \SI{1}{au}} & $ 0.03 $\\
			\hline
		\end{tabular}%
	}
	\tablecomments{For free parameters, the underscored values are fiducial values}
	\label{tab:disk_inicons}
\end{table}

\begin{table}
	\centering
	\caption{Initial conditions of planets}
	\resizebox{\linewidth}{!}{%
		\begin{tabular}{llc}
			\hline
			\hline
			{\textbf{Notation}} & {\textbf{Meaning}} & {\textbf{Value}} \\
			\hline
			$ N $&{Number of planets} & $ 3 $\\ 
			$ a_1,a_2,a_3 $& {Semi-major axis}&$ 13.2,32.3,73.7 \si{au} $\\
			$ e_1, e_2 , e_3 $& {Eccentricity}& $ 10^{-7} $ \\
			$ i_1, i_2 , i_3 $& {Inclination}& $ 0 $ \\
			$ M_1$& {Mass of Planet 1} &  $ 1.4(\alpha/\num{e-3})^{1/2} \si{M_J} $\\ 
			$ M_2$& {Mass of Planet 2} &  $ 0.2(\alpha/\num{e-3})^{1/2} \si{M_J} $\\ 
			$ M_3$& {Mass of Planet 3} &  $ 0.5(\alpha/\num{e-3})^{1/2} \si{M_J} $\\
			\hline
		\end{tabular}%
	}
	\label{tab:planet_inicons}
\end{table}
We evolve each system numerically including the planet-disk
interaction and mass accretion on the basis of the public $N$-body code
\texttt{REBOUND} \citep{Rein2011} and its extension \texttt{REBOUNDx}.
We choose an adaptive time-step integrator integrator \texttt{ias15}
\citep{Rein2014}. Within the framework of \texttt{REBOUNDx}, we implement additional
forces and modify the disk mass according to the surface density.

Our simulation starts from initial conditions following the observed
structure of the HL Tau disk (see section \ref{sec:hltau}). Table
\ref{tab:disk_inicons} summarizes the disk initial conditions that we
adopted. We take $ \SI{0.105}{M_\odot} $ as the disk mass from
\cite{Kwon2015} and set the inner edge and outer edge to be
\SI{10}{au} and \SI{80}{au}. For simplicity, we assume the stellar
mass of HL Tau is $ \SI{1}{M_\odot} $.

We use a simple power law model to characterise the vertical geometry
of the HL Tau disk. The aspect ratio is expressed as
\begin{equation}\label{aspect_ratio}
  \frac{h}{R} = \left(\frac{h}{R}\right)_{\SI{1}{au}}
  \left(\frac{R}{\SI{1}{au}}\right)^f,
\end{equation}
where $ f $ is the flaring index. We consider five flaring indices
from \num{0.15} to \num{0.35}. The reference aspect ratio at
\SI{1}{au} is fixed to be \num{0.03}, which is roughly consistent with
the values used in previous studies
\citep[e.g.][]{Dipierro2015,Kanagawa2015}.

We assume that the disk viscosity $ \alpha $ does not vary with time
and position. Considering the low turbulence level suggested by the
\rev{dust settling,} we follow \cite{Pinte2015} and take the fiducial $
\alpha $ value to be \num{3e-4}. We also consider $\alpha=$
\num{2e-4}, \num{4e-4}, \num{5e-4}, and \num{6e-4} so as to see the
dependence on $\alpha$.

We normalise the initial global accretion rate $\dot{M}_{\rm
  glob}^{\rm ini} $ using the initial disk mass calculated from the
unperturbed surface density profile(equation \ref{unp_disk_prof})
\begin{align}\label{Mdot_normalisation1}
M_{\rm disk}^{\rm ini} &= \int_{R_{\rm in}}^{R_{\rm out}} \Sigma\cdot2\pi r dr \notag\\
& = \int_{R_{\rm in}}^{R_{\rm out}}\frac{\dot{M}_{\rm glob}^{\rm ini}}{3\pi\nu}\cdot2\pi r dr.
\end{align}
Therefore,
\begin{equation}\label{Mdot_normalisation2}
\rev{\dot{M}^{\rm ini}_{\rm glob} = \frac{3}{2} \nu_{\rm 1au} \left(\frac{M^{\rm ini}_{\rm disk}}{\SI{1}{au^2}}\right) \left[\frac{3/2-2f}{\left(\frac{R_{\rm out}}{\SI{1}{au}}\right)^{3/2-2f}-\left(\frac{R_{\rm in}}{\SI{1}{au}}\right)^{3/2-2f}}\right],}
\end{equation}
\rev{where $ \nu_{\rm 1au} = \alpha\left(\frac{h}{R}\right)^2_{\rm 1au}\Omega_{K,\SI{1}{au}} $.} Note that in the present simulation, we adopt $ M_{\rm disk}^{\rm ini}  = 0.105M_{\odot}$, which is the initial mass of the disk within the range from $R_{\rm in} = \SI{10}{au}$ to $R_{\rm out}=\SI{80}{au}$. In the fiducial case, $ \dot{M}_{\rm glob}^{\rm ini} =\SI{3.82e-9}{M_\odot yr^{-1}} $.

Table \ref{tab:planet_inicons} summarises the planet-related initial
conditions. \citetalias{Simbulan2017} interpreted the HL Tau disk
substructure as the existence of four or five planets. Since
hydrodynamic simulations \citep{Dipierro2015, Jin2016} have shown
that the major substructures of the HL tau disk can be well reproduced
by three planets, we consider three planets
initially located at \SI{13.2}{au}, \SI{32.3}{au} and \SI{73.7}{au},
corresponding to the first, second and fourth planets in
\citetalias{Simbulan2017}. \rev{For reference, the mid-plane disk temperature at the location of each planet is \SI{45}{K}, \SI{29}{K} and \SI{19}{K} from inner to outer, if the temperature at \SI{13.2}{au} gap is taken as \SI{45}{K} (Table 2, \cite{Pinte2015}) and the flaring index is the fiducial value \num{0.25}.}

We assume that the planets are co-planar and initially in
near-circular orbits by setting the initial eccentricity to be $
\num{e-7} $ for all the three planets. The adopted planetary masses
are given from the width of the gap to the planetary mass using an
empirical formula following \citet{Kanagawa2016}.  We choose the
planetary masses to be $ \SI{0.77}{M_J} $, $ \SI{0.11}{M_J} $, and $
\SI{0.27}{M_J} $ for our fiducial model of $\alpha=$ \num{3e-4}, and
scale them by $ \propto \alpha^{1/2}$ (See Equation (5) of \cite{Kanagawa2015} and also Table
\ref{tab:planet_inicons}).

\section{Results}
\label{sec:rslt}

\begin{figure*}
	\centering
	\includegraphics[width=0.8\linewidth]{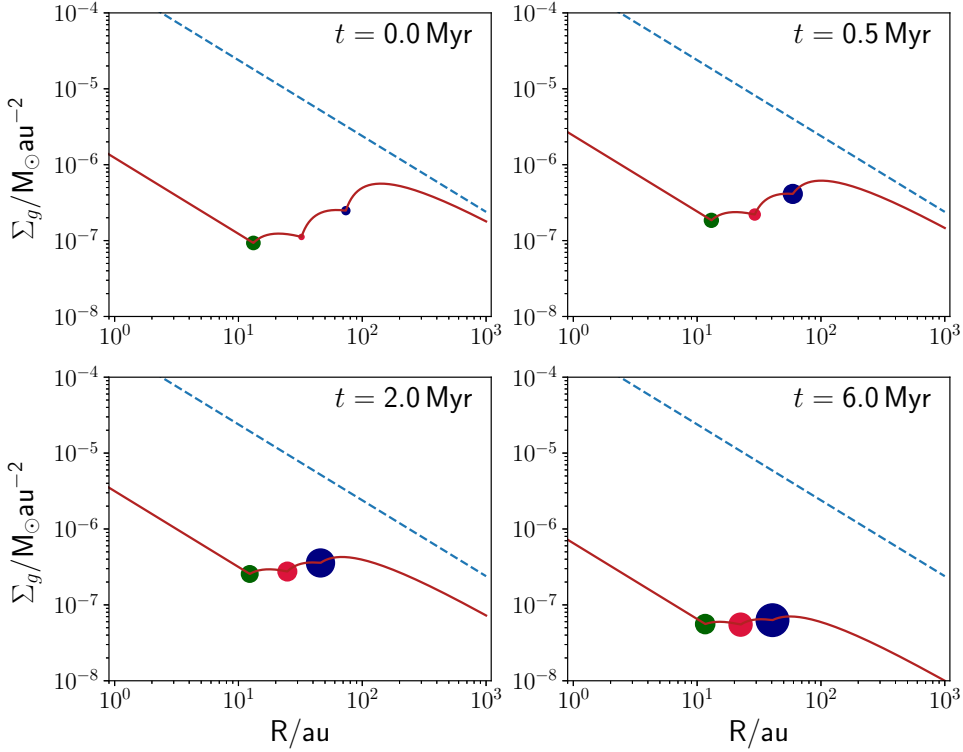}
	\caption{Disk surface density profile at four different epochs. Each coloured dot indicates the respective position of the planet, where the size of the dot is proportional to the mass of the planet. As a reference, the blue dotted line is the disk profile without the planets at t = 0.}
	\label{fig:surface_den_visual}
\end{figure*}

\subsection{Results before the dispersal of the disk}

As shown in Table \ref{tab:disk_inicons}, we perform $5\times3\times5=75$ simulation runs by varying disk parameters.  The simulation stops at $t=3\tau_{\rm disk}$ when the gas component of the disk is sufficiently removed from the system.

Figure \ref{fig:amet_evo} shows an example of planetary evolution with fiducial parameters: flaring index $ f = 0.25 $, disk lifetime $ \tau_{\rm disk} = \SI{2}{Myr} $ and viscosity $ \alpha = \num{3e-4} $. In $ \SI{6}{Myr} $, orbits of planets $ 1,2,3 $ shrink from $ (13.2,32.3,73.7)\si{au} $ to $ (11.6,22.5,40.7)\si{au} $, respectively, which remain widely separated. Their masses increase from $ (0.77,0.11,0.27)\si{M_J} $ to $(1.6,2.3,4.6)\si{M_J} $. The relation between semi-major axis and mass agrees with the result of \cite{Tanaka2019} who have investigated the evolution of a single planet within the disk.

Due to the large initial mass and the low surface density, the innermost planet (Planet 1) barely migrates throughout the disk stage. The outer two planets (Planet 2 and 3) quickly grow and migrate inward for the first \SI{0.5}{Myr}. As the mass of planets keep growing, the migration of the two planets gradually slows down, and they eventually approach their final positions as the disk surface density decays.

Over the entire migration, eccentricities of the three planets are very small since the slow Type II migration does not lead to any close encounter between the planets. The eccentricities of all three planets are below \num{0.01} for $t<\SI{2}{Gyr}$. For Planet 2, there is a slight excitation of eccentricity at around \SI{3.6}{Myr} due to the approach of Planet 3. Even after the excitation, eccentricities of Planet 2 and 3 fluctuate around \num{0.025} and \num{0.015} and nearly stop growing.

The migration time-scales in Figure \ref{fig:amet_evo} indicate that the outer planet initially migrates faster than the inner planet. In the first \si{Myr}, the migration time-scales of Planet 2 and Planet 3 gradually increase, as their gaps become deeper due to the mass growth. The accretion rates of the outer planets also drop, so more mass can flow inside and boost the migration of Planet 1, which explains the decrease of the migration time-scale of Planet 1.  The migration time-scales of all the three planets exceed \SI{10}{Myr} at $ t = \SI{2}{Myr}$. They eventually exceed \SI{100}{Myr} at $t = \SI{6}{Myr}$, and their migration ceases practically.

\begin{figure}
\centering
\includegraphics[width=\linewidth]{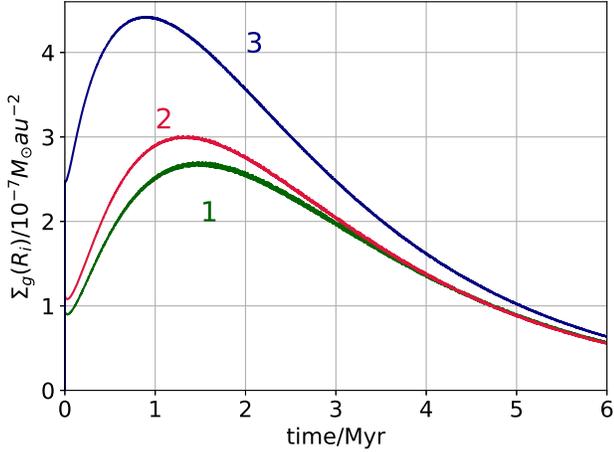}
\caption{Evolution of the disk surface density at the location of each
  planet}
\label{fig:planet_sufden_evo}
\end{figure}

Figure \ref{fig:surface_den_visual} plots the global surface density profile (\textit{cf.} equation \ref{multip_sigmaunp}) at four different epochs of the fiducial run, while Figure \ref{fig:planet_sufden_evo} is the evolution of the surface density evaluated at the location of each planet. At $t = 0$, the strong mass accretion onto planets 2 and 3 creates steep bumps on the density profile, and the surface density around Planet 1 is quenched to less than \SI{1}{\percent} of the unperturbed value.  As Planets 2 and 3 continue to grow, their accretion rates drop as $ \dot{M}_i \propto M_i^{-2/3} $. Thus the mass flows through the outer planets and accretes inward preferentially on planet 1 and the star.  At $ t = \SI{2}{Myr} $, the planetary system becomes more closely packed and accretion rates of all the planets drop. As a result, density bumps around the planets almost vanish, while the density at the vicinity of each planet reaches the maximum at this epoch. At $ t = \SI{6}{Myr} $, the density bumps become further negligible and the surface densities at three planets become nearly identical.

\begin{figure}
	\centering
	\includegraphics[width=0.9\linewidth]{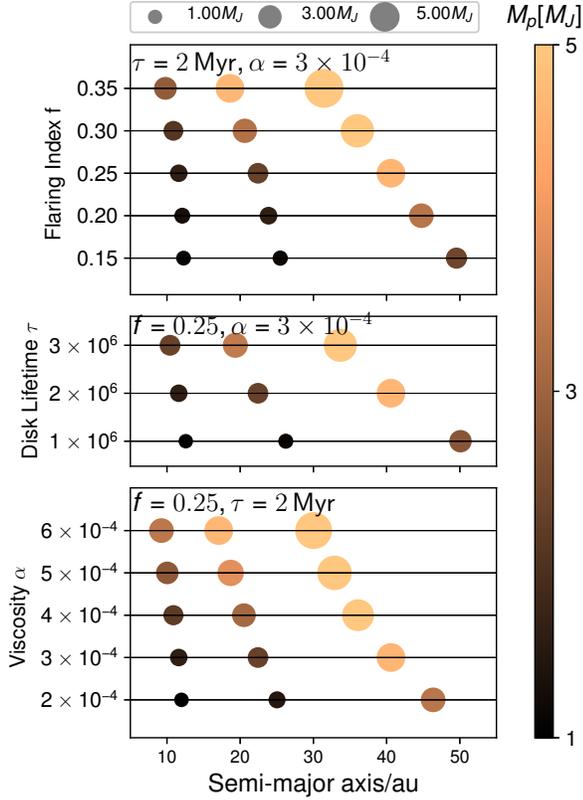}
	\caption{Planetary configurations at $ t = 3\tau_{\rm disk} $
          with different disk parameters. The size of the marker is
          proportional to the mass of the planet.}
	\label{fig:planet_config}
\end{figure}

Figure \ref{fig:planet_config} illustrates the configurations of the planetary systems at $3\tau_{\rm disk}$ with different disk parameters. For each plot, we vary only one free parameter while the other two are fixed as their fiducial values. Overall the mass of the outer planet is larger than that of inner one, because the outer planet grows \rev{faster} than the inner one. The gas accretion onto the outer planet decreases the gas surface density around the inner planets. Because of the lower gas density, the migration and mass growth of the inner ones are suppressed until the gas accretion onto the outer planet becomes small due to the formation of the deep gap. This is why the evolution of the inner planet is less sensitive to the disk parameters compared to the outer planet. The final semi-major axis and planetary mass change monotonically with the value of each parameter.

When the flaring index $f$ increases from \num{0.15} to \num{0.35}, the final semi-major axis decreases and final mass increases. \rev{As the flaring index become larger, the aspect ratio of the disk increases quickly with radii, resulting in a much shallower gap since the gap depth is sensitive to the aspect ratio to the fifth power (see equations \ref{6.1siggap} and \ref{eq:def-K}). While the ratio of Lindblad torque decreases with the flaring index since $ T \propto R^{-(1-2f)}$ and $ \Sigma \propto R^{-(0.5+2f)} $ (see equation~\ref{eqn:gammal14}), the effects caused by a shallower gap dominates both the migration time-scale (see equation~\ref{taua_approx}) and accretion rate. Consequently, the planets experience stronger migration and accretion, and eventually move further inward and become more massive. This effect is stronger at large radii, and hence the outermost planet is most affected, as shown by Figure \ref{fig:planet_config}.}

When $ \tau_{disk} $ increases from \SI{1}{Myr} to \SI{3}{Myr}, the final semi-major axis decreases and final mass increases monotonically. Since $ \tau_{\rm disk}$ is basically a disk lifetime, the above trend is easily understood as planets in a long-lived disk simply have longer time for migration and accretion.

Finally, a higher viscosity implies that planets experience stronger migration and accretion, as evidenced by the decreasing semi-major axis and increasing final mass. Physically, it is because the angular momentum transfer becomes more efficient with a higher viscosity.  A more quantitative discussion can be made from equation (\ref{eq:def-K}), \rev{which is similar to the explanation of flaring index dependence.} Since the diffusion time-scale is proportional to $\nu^{-1}\propto \alpha^{-1}$, a larger value of $ \alpha $ leads to a faster gas diffusion, resulting in a shallower gap carved by a planet with the same given mass, {\it i.e.}, the $K$ value is smaller. Thus the migration time-scale is shorter and migration is faster. The mass accretion dependence on $K$ is understood similarly: the accretion density $ \Sigma_{acc} $ is higher for a shallower gap, and therefore a planet grows factor with a higher accretion rate and becomes more massive.

We find that 74 out of our \num{75} runs remain stable until the end of the disk gas removal at $3\tau_{\rm disk}$, but the remaining run with $ f = 0.35 $, $ \tau_{\rm disk} = \SI{3}{Myr} $, and $ \alpha = \num{6e-4}$ turns out to be unstable before the epoch. In this case, the innermost Planet 1 is ejected due to gravitational scattering, and the remaining Planet 2 and 3 are in fairly eccentric orbits($ e_2 = 0.55, e_3 = 0.21 $). We therefore exclude this single unstable case in our follow-up analysis. \rev{Nevertheless, this unstable case stresses the importance of the planet-planet interaction even before the dispersal of the disk, which is absorbed in our N-body approach. When planets are close to each other, particularly inside the resonance overlapping region, planet-planet interactions can couple with the planet-disk interactions and play an important role in shaping the stability of the configuration.}

\subsection{Results at $ \SI{10}{Gyr} $ after disk dispersal}

\begin{figure*}
	\centering
	\includegraphics[width=\linewidth]{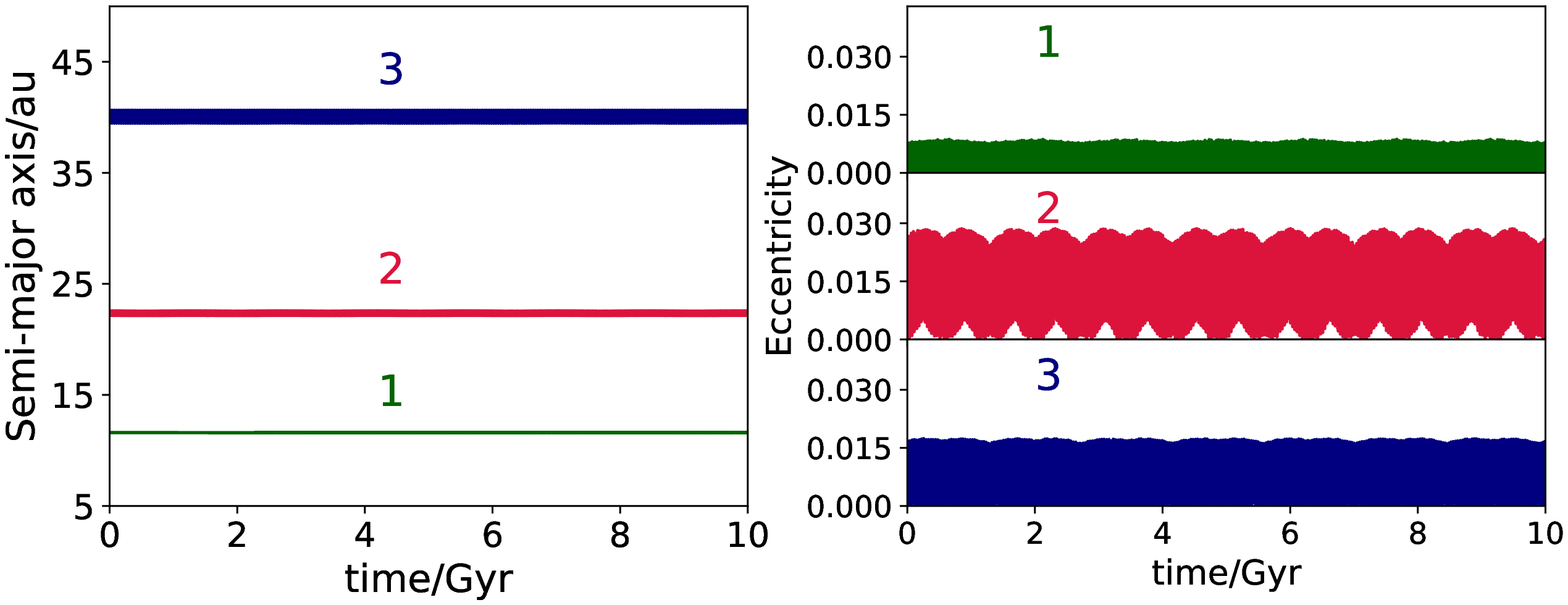}
	\caption{Evolution of the semi-major axis and eccentricity of the fiducial case \SI{10}{Gyr} after disk dispersal in our fiducial case (the parameters are the same as these in the case shown in Figure~\ref{fig:amet_evo}).}
	\label{fig:evo_after_dispersal}
\end{figure*}

We continue evolving the \num{74} systems from the final planetary configurations at $3\tau_{\rm disk}$ of the previous disk stage using purely $N$-body simulation after the disk dispersal. When one of the planets is ejected from the system or collide with star/other planet, we stop the simulation and mark the run unstable. Otherwise the system is integrated up to \SI{10}{Gyr}. \rev{The simulation runs are effectively 2D simulation since no inclination has been introduced throughout our simulations.}

Figure~\ref{fig:evo_after_dispersal} shows the orbital evolution of the planets in our fiducial case after the disk dispersal. Clearly there is no significant change in the semi-major axis and eccentricity of the planet during \SI{10}{Gyr}. We also confirmed that the majority of the systems (\num{69} out of \num{74}) remain stable until \SI{10}{Gyr} as in the case shown in Figure~\ref{fig:evo_after_dispersal}.

A low fraction of unstable systems may imply that most of the planetary configuration achieved through physical disk-planet evolution is indeed stable.  This is in contrast to the fact that a significant fraction of numerical simulations exhibit unstable outcomes, most likely due to their relatively artificial initial conditions.  Our result indicates that the production of misaligned planets and Hot Jupiters via the instability channel is rather inefficient, at least for the HL Tau disk.

The remaining \num{5} systems become unstable within \SI{0.004}{Gyr} to \SI{5.7}{Gyr}. We find that the instability time is extremely sensitive to the numerical treatment because of the chaotic nature of such systems, and even a tiny numerical truncation error could result in a \SI{50}{\percent} change of the instability time. The precise value of the instability time can be trusted roughly within an order of magnitude. More detailed discussions on the instability is given in section \ref{sec:consistency} below.

\section{Discussion}
\label{sec:discussion}

\subsection{Consistency with the previous stability criteria}
\label{sec:consistency}

\begin{figure}
	\centering \includegraphics[width=\linewidth]{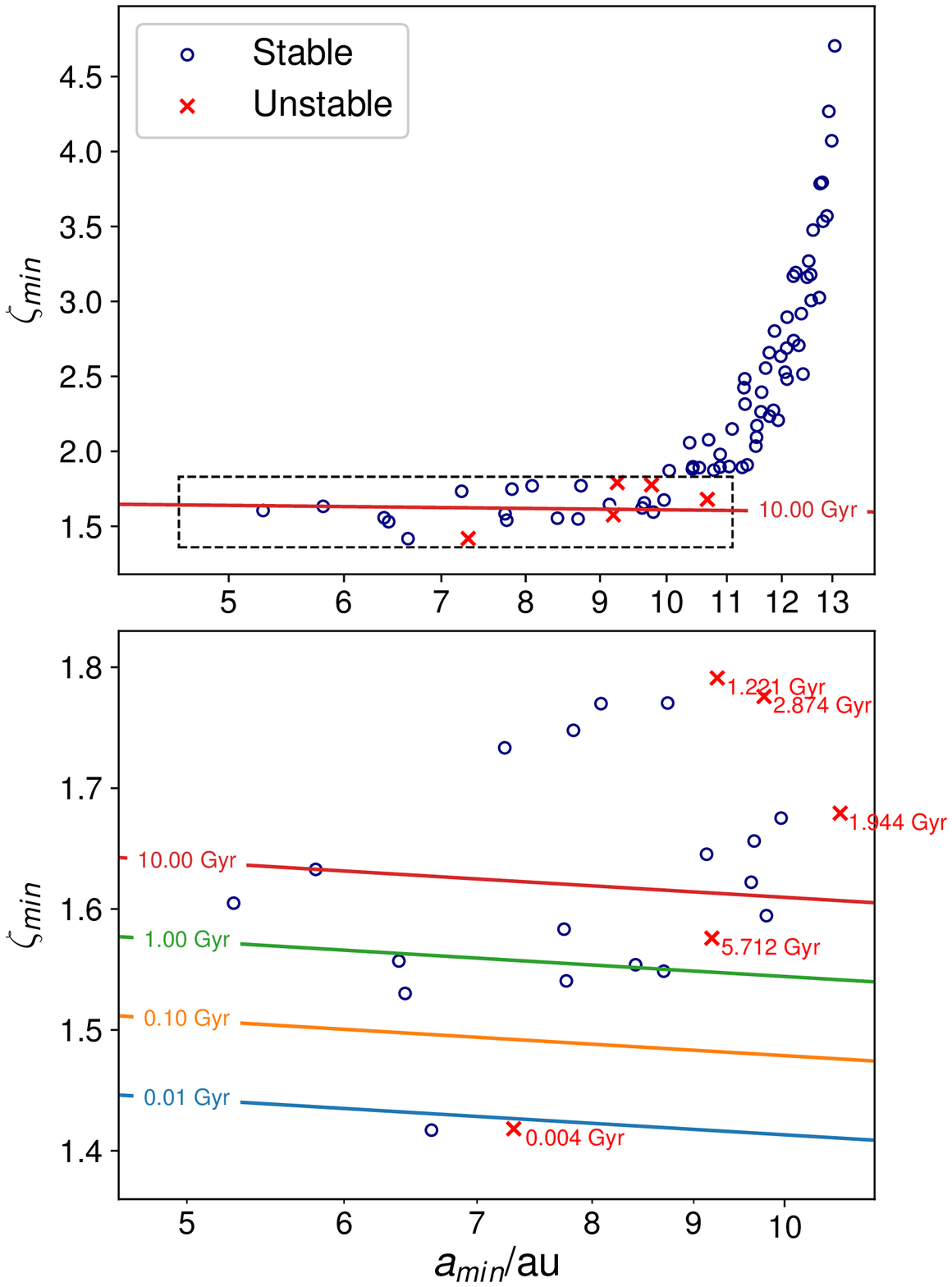}
	\caption{$ \zeta_{\rm min} $
          against the semi-major
          axis of the innermost planet. The figure below is the
          enlarged dotted area on figure above. Unstable cases and
          corresponding instability time are tagged in red. The
          coloured lines are the instability time contours based on
          the result of \cite{Morrison2016}.}
	\label{fig:Stability}
\end{figure}
The stability criterion of a multi-planetary system has been extensively investigated. In particular, a number of previous work \citep[e.g.][]{Gladman1993,Chambers1996a,Marzari2002,Quillen2011,Tamayo2016,Wu2018} pointed out on the basis of the Hill stability argument that a mutual orbital separation of planets plays an important role in their long-term stability. However, when the planetary mass is large, the mean resonance becomes important and those empirical relations based on mutual Hill radius are not directly applicable. We found that the instability time predictions given by \cite{Chambers1996a} and \cite{Marzari2002} significantly underestimate the lifetime of our simulated systems. \rev{For instance, the criterion given by \cite{Marzari2002} predicts half of the systems to be unstable, though most of the systems are stable in our simulations. This discrepancy is originated from the planetary mass dependence of the stability criterion.}

\cite{Morrison2016} performed a series of simulations with three and five equal-mass planets and found that the mean resonance overlap is a better measure of the orbital stability in the high mass regime larger than Jupiter. On the basis of their argument, we consider the following empirical log-linear relation between an orbital instability time $T_{\rm ins}$ and a dimensionless orbital spacing $\zeta$:
\begin{gather}
\label{inst_time_mk}
\log(T_{\rm ins}/P_{\rm inner}) = c\zeta + d, \\
\label{zeta}
\zeta_{\rm min} = \min_{i=0,1,2} \zeta_i, \\
\label{zeta-i}
\zeta_i \equiv  \frac{a_{i+1}-a_i}{\delta a_{{\rm ro},i}}\Big|_{\rm init}, \\
\label{eq:delta-a}
\delta a_{{\rm ro},i}  \equiv 1.5\left(\frac{M_p}{M_*}\right)^{2/7}a_i ,
\end{gather}
where $P_{\rm inner}$ is the orbital period of the inner-most planet, $M_p$ is the planetary mass, $\delta a_{{\rm ro},i}$ is the separation where two first-order mean resonances overlap, and $c$ and $d$ are numerical constants.

Note that the initial conditions of the simulations of \citet{Morrison2016} are such that $\zeta_i$ are independent of $i$ for a given system, and thus $\zeta_{\rm min}=\zeta_i$. Then we fit equation (\ref{inst_time_mk}) to their result (left panel of their Figure 3 with three equal-mass planets), and found that $c \approx 15.3$ and $d \approx -16.1$ for planets more massive than $ \SI{e-3}{M_\odot} $ reproduce their data roughly within one order-of-magnitude.

Since our systems consist of three unequal-mass planets, and $\zeta_i$ depend on $i$, we cannot directly adopt equation (\ref{inst_time_mk}) for the stability condition. \cite{Deck2013} found the same relation hold if the mass ratio in the original $\delta a_{{\rm ro},i}$ is replaced by the sum of the mass ratios of two planets, \textit{i.e.}, $ \delta a_{{\rm ro},i}^\prime $. We compared both scaling factors and found $ \delta a_{{\rm ro},i}^\prime $ is more consistent with our results with unequal planetary mass. Thus we extrapolate  equation (\ref{inst_time_mk}) and rewrite it in terms of $\delta a_{{\rm ro},i}^\prime$ as follows:
\begin{gather}
\log\left(\frac{T_{\rm ins}}{\si{yr}}\right) = c\zeta^\prime_{\rm min} + d
+ \frac{3}{2} \log \left(\frac{a_{\rm inner}}{\SI{1}{au}}\right) \notag\\
- \frac{1}{2} \log \left(\frac{M_*}{\si{M_\odot}}\right).
\label{eq:Textrapolated}
\end{gather}
In the above expression, $a_{\rm inner}$ is the semi-major axis of the innermost planet
evaluated at the initial epoch for the $N$-body simulation ({\it i.e.,} at the end of the disk-planet interaction run), and
\begin{gather}
  \zeta^\prime_{\rm min} = \min_{i=0,1,2} \zeta_i^\prime, \\
  \zeta_i^\prime \equiv \frac{a_{i+1}-a_i}{\delta a_{{\rm
        ro},i}^\prime} \Big|_{\rm init},\\
  \delta a_{{\rm ro},i}^\prime
\equiv 1.5\left(\frac{M_i+M_{i+1}}{M_*}\right)^{2/7}a_i.
\end{gather}
Figure \ref{fig:Stability} plots $ \zeta_{min} $ against the $a_{\rm inner}$ for 74 runs. Circles indicate 69 systems that are stable until $ t= \SI{10}{Gyr} $, while the remaining 5 systems (crosses) become unstable within the integration time.  As a reference, equation (\ref{eq:Textrapolated}) predicts that $62$ systems are stable, and our simulation confirmed that $ 59 $ out of the $ 62 $ systems ($95.2\%$) are indeed stable. On the other hand, among the \num{12} systems that are predicted to be unstable within \SI{10}{Gyr}, only \num{2} systems become unstable, indicating that equation (\ref{eq:Textrapolated}) systematically underestimates the instability time to some extent. \rev{Although the actual instability time can deviate up to two or three orders of magnitude from the predicted line, equation (\ref{eq:Textrapolated}) outperforms other criteria and is roughly consistent with our result that most of systems are stable within \SI{10}{Gyr}.}

\subsection{Period ratios and mean motion resonance}

\begin{figure}
	\centering
	\includegraphics[width=\linewidth]{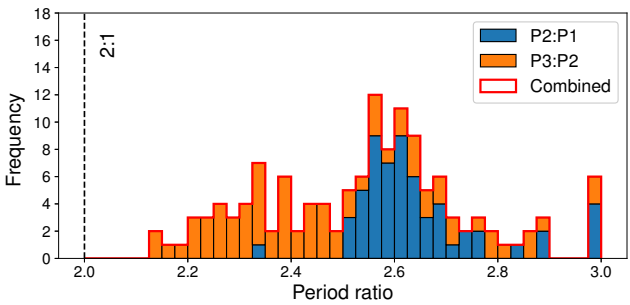}
	\caption{Histogram of period ratios of adjacent planet pairs
          at $ t = 3\tau_{disk} $. The bin size is \num{0.025}.}
	\label{fig:hist_pratio}
\end{figure}

The stability of systems emerging from the HL Tau disk may be understood as well in the context of the mean-motion resonance (MMR) capture during a convergent migration \citep{Mustill2011,Goldreich2014,Deck2015,Tamayo2017}. \cite{Obertas2017} found that the stability of a planetary system is enhanced if a pair of planets are in a near-resonant state. \cite{Tamayo2017} performed simulations for the TRAPPIST-1 planetary system, and concluded that the disk migration can produce a resonant chain of planets, which significantly stabilises the system.

Motivated by these studies, we plot the histogram of period ratios between adjacent planets, $P_2/P_1$ and $P_3/P_2$, at the initial epoch (Figure \ref{fig:hist_pratio}).  The period ratios fall within the range from \num{2.1} to \num{3.0}.  \rev{The minimum period ratio is clearly above \num{2.0},} indicating that no planet pair has ever entered the strongest first-order resonance region. The cut-off implies the existence of a strong co-migration between two planets; when the outer planet approaches the inner planet from outside of 2:1 period ratio, the inner planet is forced to migrate together, which prevents the planet pair from further approaching and entering the first-order MMR region.

The period ratios of the two adjacent pairs also exhibit different statistical distributions. The period ratios of the outer pair ($P_3/P_2$) are widely distributed between \num{2.1} and \num{2.9} range, while those of the inner pair ($P_2/P_1$) are in a narrow range between \num{2.5} and \num{2.8} with a peak at \num{2.6}. These distributions indicate that the outer pairs are more closely-packed than the inner pairs on average. One possible explanation is that the period ratio is determined by the migration speed difference of the pair. In the case of the outer pair, the faster migration speed of Planet 3 allows Planet 3 to reach the region closer to the 2:1 resonance before the co-migration starts. In the case of the inner pair, however, the migration speed of Planet 2 slows down and their orbital separation is wider before Planet 2 approaches Planet 1.

\begin{figure}
	\centering
	\includegraphics[width=\linewidth]{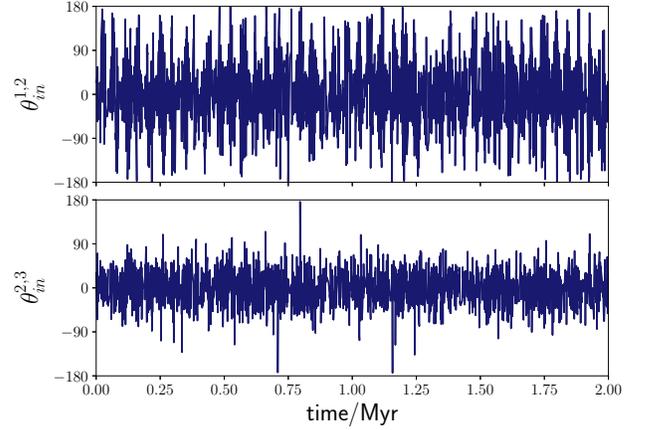}
	\caption{2:1 resonant argument evolution of the inner planet in planet-pair 1,2 and planet-pair 2,3 of the fiducial case after disk dispersal.}
	\label{fig:10}
\end{figure}

In order to see the depth of the resonance, we plot in Figure
\ref{fig:10} the 2:1 resonant argument of the inner planet in the
fiducial run; upper and lower panels correspond to the inner and outer
planet-pairs, respectively. The 2:1 resonant argument of the inner
planet is defined as
\begin{equation}\label{resonant_arg}
\theta_{in} = \lambda_{in} - 2\lambda_{out} + \varpi_{in},
\end{equation}
where $ \lambda $ is the mean longitude and $\varpi$ is the longitude of periastron with subscripts ``in" and ``out" denoting the inner and outer planet, respectively.

The resonant argument of Planet 1 in the inner planet-pair shows significant fluctuation around $ \theta = \SI{0}{\degree} $ with an amplitude of nearly \SI{180}{\degree}. The resonant argument of Planet 2 in the outer planet-pair shows a much `well-behaved' oscillation at an amplitude of \SI{90}{\degree}, with sporadic spikes almost touching $\pm \SI{180}{\degree} $.  It is expected that the outer pair of planets are in a deeper resonance state because they are closer to each other than the inner pair, shown by the period ratio histograms. Both of the resonant arguments indicate that the planets are neither in deep resonance nor totally out-of-resonance, since both arguments are not librating between $\pm \SI{180}{\degree} $ throughout the time. It is not clear that how much such a weak resonance state can contribute to the long-term stability. However, as already pointed out by dynamical stability studies concerning the system HR 8799 (see section \ref{sec:implications}), being in deep resonance state is not a necessary condition for long-term stability.

\subsection{Implications for widely-separated massive planetary systems like HR 8799 and PDS 70}
\label{sec:implications}
\rev{Our simulation shows that a stable, widely-separated super-Jupiter system is the dominant outcome of the three-planet configuration that we extracted from the current HL Tau observation, within the parameter space that we explored.} Such a configuration is beyond the detection range of the previous prolific planet-hunting telescopes like \textit{Kepler}, and therefore our result is largely different from the majority of multi-planetary systems known to-date, which are generally closely-packed with earth-size planets. There are a couple of such systems, HR 8799 and PDS 70, which have been detected so far via direct imaging or high-resolution spectroscopy.

Both HR 8799 and PDS 70 host planets several times massive than Jupiter separated by a large distance.  The HR 8799 system is known as the first multi-planetary system discovered via direct imaging \citep{Marois2008,Marois2010}. It consists of four giant planets located at \SI{16.4}{au}, \SI{27}{au}, \SI{42.9}{au} and \SI{68}{au}, with estimated mass ranging from $ \SI{7}{M_J} $ to $\SI{10}{M_J} $ \citep{Marois2010, Soummer2011}. The best-fit model shows that the orbits are near co-planar with a low eccentricity $ \sim 0.1 $. PDS 70 hosts two planets with semi-major axes \SI{20.6}{au} and \SI{34.5}{au} and planetary masses estimated to be $ \SI{7}{M_J} $ to $ \SI{8}{M_J} $ \citep{Keppler2018,Haffert2019}. Results of our simulations show the mass of the outer planet is likely to be larger than that of the inner planet. Moreover, the period ratios of the planet pair can be close to but outside the 2:1 mean-motion resonance. The systems of HR~8799 and PDS 70 satisfy the above features, which may indicate that these planets may be formed in protoplanetary disks similar to HL~Tau.

In terms of the planet formation picture, PDS 70 agrees with what we assume for HL Tau. Its two planets are observed inside a gap of the transiting disk, and the H$ \alpha $ emission indicates that gas accretion onto the planets is still ongoing \citep{Keppler2018}. The formation channel of HR 8799, however, remains to be understood. Previous studies \citep[e.g.][]{Boss2011,Vorobyov2013} suggest that planets in HR 8799 are likely to be produced via {\it in-situ} gravitational fragmentation, due to the difficulty to apply the standard core accretion model at such a large distance. Our simulation shows it is physically possible to form such wide-separated systems similar to HR 8799 via migration and accretion, if our planetary interpretation on the current substructure of the HL Tau disk is adopted. \rev{However, such a initial condition bypasses the fundamental difficulty of forming planets at large distances from the star, and may only be an intermediate state in a larger evolution picture.}

The dynamical stability of HR 8799 remains as yet another unsolved question. \cite{Fabrycky2010} pointed out that HR 8799 may have an instability time much shorter than the stellar age, and some sort of resonance is required to stabilise the system.  \cite{Gozdziewski2018} proposed that the stable configuration of HR 8799 may be attributed to the resonance capture resulting from the convergent migration. In a different context, our simulation result confirmed that the convergent migration is a natural solution to the long-term stable configuration, even though our systems are in marginal resonance. In fact, our results are similar to the results shown by Figure 10 in \cite{Gotberg2016}, who found the presence of systems only `sporadically' in resonance without being locked (See Figure \ref{fig:10}).

Due to the limited number of such observed systems, we are unable to conclude anything definite at this point. Since we have seen some interesting connections with the two observed examples, however, we may speculate that those widely-separated planetary systems originate from the HL Tau like disks may commonly exist. This speculation can be tested further by future possible detection of systems similar to HR 8799 and PDS 70 using the current and next-generation telescopes, such as \textit{WFIRST} and \textit{JWST}.

\subsection{Caveats and future work}
We would like to mention a few issues that we need to address in the future. First, we apply a migration and accretion model that are originally developed for a single planet case to a three-planet system assuming that the gaps around those planets can be treated separately. If two massive planets are sufficiently close, however, their gaps may merge into a single common gap \citep[e.g.,][]{Duffell2015a,Cimerman2018}. If that is the case, our approximation does not hold, and we need to improve the model.  Quantitatively it is not yet clear how two giant planets behave in a common gap, and we have to perform hydrodynamic simulation to understand the behaviour.

We assume that the $\alpha $ viscosity does not vary with time and position for simplicity, but it may not be a good approximation in some cases. It is widely believed that the gas accretion in the disk is caused by the turbulent viscosity driven by MRI \citep[][for a review]{Turner_PPVI}. However, the onset of MRI is related to non-ideal MHD effects, including Ohmic dissipation, Hall effect, and ambipolar diffusion. For instance, since the MRI turbulence is suppressed due to Ohmic dissipation, the turbulent viscosity in the mid-plane can be small at intermediate radii of the protoplanetary disk, so-called `dead-zone' \citep[e.g.,][]{Gammie1996}. In such a region, the value of $\alpha$ should be smaller than that in the other region. Moreover, recent studies have shown that the angular momentum can be transported by the magnetically driven wind \citep[e.g.,][]{Suzuki2009,Bai2016,Suzuki2016}.

The above effects may change the disk evolution and affect the planetary evolution. We also assume that the disk structure instantaneously reaches that in steady state. However, this assumption is not always valid, especially in the cases with a low viscosity. Discussion on those effects is beyond the scope of the present paper, and intensive numerical work is required that solves the mass and angular momentum transfer equations including the magnetic field simultaneously.

We consider three planets in the HL tau that correspond to the major gaps, following several recent results of hydrodynamic simulations. However, these hydrodynamic simulations do not rule out the presence of small planets that do not form any gap, nor additional planet formation in the later stage. Increasing the number of planets would generally destabilise the system \citep[e.g.,][]{Chambers1996a,Marzari2002,Wu2018}, and we plan to vary the number and mass of planets to see their effect on the stability of the systems. 

\rev{Our current simulations only consider co-planar cases, which means the systems are effectively 2D. Introducing inclinations can affect the stability in both ways. On the one hand, inclinations allow additional mechanism, such as Lidov-Kozai oscillations (\cite{Kozai1962}) to act in, which may cause orbit crossing and destabilise the system. On the other hand, large inclinations can also prevent close encounters of planets, therefore stabilise the system \citep[e.g.][]{Wang2019}. We plan to include inclinations and investigate its effects on the dynamical stability in the next step.}

Finally, the current study focuses on the HL Tau disk alone, because it is the first observed system that exhibits significant substructure, and thus has been well-studied.  Now a number of such systems have been observed including a project of DSHARP \citep{Andrews2018} among others \citep[e.g.,][]{Dong2018, Long2018,VanderMarel2019,Tsukagoshi2019}. We plan to apply our methodology to other planetary systems to see to what extent the outcomes of planetary systems emerging from the HL Tau are generic.

\section{Summary}
\label{sec:summary}
The planet-disk interaction is a widely accepted interpretation of the ring and gap substructure in the HL Tau disk. This interpretation allows us to extract the orbital and mass information of the HL Tau planetary system as initial conditions, which can be used in predicting its evolutionary outcomes.
	
We consider three planets initially in co-planar, circular orbits with semi-major axis corresponding to the three deepest gaps in the HL Tau disk, and perform two-stage numerical simulations. In the disk stage, we include both realistic migration and accretion processes coupling with an adaptive disk profile, and evolve the system until the disk is sufficiently dispersed. After the disk dispersal, we continue to evolve the system using purely gravitational N-body simulations, and examine the orbital stability up to \SI{10}{Gyr}. Our main conclusions are summarized below:

\begin{itemize}
\item We are able to produce a variety of widely separated multi-planetary systems by varying the disk parameters from the HL Tau. When we increase the values of the flaring index, disk lifetime and the $ \alpha $ viscosity, both the migration and accretion become stronger. Both the final semi-major axis and mass of the inner planet are less sensitive to the disk parameters, while those of the outer planet are more sensitive. Although the period ratios of most planet pairs are larger than \num{2}, these pairs are in marginal 2:1 resonance. Planets of the outer pair (Planet 2 \& 3) are closer to each other than those of the inner pair (Planet 1 \& 2).

\item We found that majority of systems are stable.  We run 75 models with different disk flaring index, viscosity, and gas dispersal time-scale, and 69 out of the 75 models remain stable for 10 Gyr.  In our fiducial model, for instance, three planets migrate from $ \num{13.2}, \num{32.3}, \num{73.7}\si{au} $ to $ \num{11.6}, \num{22.5}, \num{40.7}\si{au} $ within \SI{6}{Myr}, while their masses also grow from $ \num{0.77}, \num{0.11}, \num{0.27}\si{M_J} $ to $ \num{1.6}, \num{2.3}, \num{4.6}\si{M_J}$.  The eccentricities of all planets are below \num{0.03} due to the absence of close-encounter, and the migration is convergent. 

\item The stability of those systems is roughly consistent with the previous empirical criterion by \citet{Morrison2016}. It is also explained by the lack of adjacent planet pairs crossing the 2:1 resonance zone.

\item The resulting architecture of widely-separated massive planetary systems is similar to the observed systems including HR 8799 and PDS 70, indicating an interesting link between the HL Tau like disks and the origin of such wide-orbit giant-planet systems.
\end{itemize}

Our current results imply that the HL Tau disk predicts a fairly stable architecture with longer-period planets. While such systems remain to be detected in future with longer-time monitering, there are a couple of systems resembling our prediction. It is also possible that the HL Tau disk is an exceptionally stable system and/or that four or five planets in the disk may increase the fraction of unstable systems. We are now addressing those questions, and plan to report the result in the future work.

\section*{Acknowledgement}

We thank Makiko Nagasawa, Shigeru Ida, Eiichiro Kokubo, Alessandro Alberto Trani and Hidekazu Tanaka for discussion on dynamical evolution of multi-planetary systems. The numerical simulation has been carried out using Linux clusters at Center for Computational Astrophysics, National Astronomical Observatory of Japan. We compiled the \texttt{C} code using standard \texttt{gcc} compilers, and the results were analysed in \texttt{Python} $ 3.7 $ environment. K.D.K. acknowledges the support from Research Center for the Early Universe, the University of Tokyo.  This work is supported partly by Japan Society for the Promotion of Science (JSPS) Core-to-Core Program “International Network of Planetary Sciences”, and also by JSPS KAKENHI Grant Numbers JP18H01247 and JP19H01947 (Y.S.), and JP19K14779 (K.D.K).


\appendix

\section{Disk profile hosting multiple planets}
\label{sec:appendix1}
For a static disk hosting $ N $ planets, each located at $ R_i $ with accretion rate $ \dot{M}_i $, $  i\in\{ 1,2,...,N\} $, the conservation of mass can be written as
\begin{align}
\dot{M}(R) = \begin{cases}
\dot{M}_*\quad(R<R_1),\\
\dot{M}_*+\sum\limits_{i=1}^{n}\dot{M}_i\quad(R_n<R<R_{n+1}),\\
\dot{M}_*+\sum\limits_{i=1}^{N}\dot{M}_i\quad(R>R_N),
\end{cases}
\label{mdotr}
\end{align}
where $ n\in\{1,2,...N-1\} $. Similarly, conservation of the angular momentum gives
\begin{align}
\dot{M}(R)j(R) = \begin{cases}
\dot{M}_*j_*\quad(R<R_1),\\
\dot{M}_*j_*+\sum\limits_{i=1}^{n}\dot{M}_ij_i\quad(R_n<R<R_{n+1}),\\
\dot{M}_*j_*+\sum\limits_{i=1}^{N}\dot{M}_ij_i\quad(R>R_N),
\end{cases}
\end{align}
where $ j_i = j(R_i) = R_i^2\Omega_i $ is the specific angular momentum. When $ R_n<R<R_{n+1} $, we assume the zero-torque boundary conditions at $ R = R_i $, and solve mass transfer equation and obtain
\begin{align}
\label{mtrans_eqn}
3\pi\nu j(R)\Sigma_g(R) = \dot{M}(R)j(R) -\dot{M}_*j_*-\sum\limits_{i}^{n}\dot{M}_ij_i.
\end{align}
where $ \Sigma_g(R) $ is the gas surface density at distance $ R $, $
\nu $ is the kinematic viscosity that can be expressed by the$
\alpha $ parameter \citep{Shakura1973}. Substituting
equation (\ref{mdotr}) to equation (\ref{mtrans_eqn}), we have
\begin{align}
\Sigma_g(R) &=
\frac{\dot{M}_*}{3\pi\nu}\left(1-\frac{j_*}{j(R)}\right) +
\sum\limits_{i=1}^n\frac{\dot{M}_i}{3\pi\nu}\left(1-\frac{j_i}{j(R)}\right)\notag\\ &=\frac{\dot{M}_*}{3\pi\nu}\left(1-\sqrt{\frac{R_*}{R}}\right)
+
\sum\limits_{i=1}^n\frac{\dot{M}_i}{3\pi\nu}\left(1-\sqrt{\frac{R_i}{R}}\right).
\label{6.1sigmaunp}
\end{align}
This expression is what we use in the main text, equation (\ref{multip_sigmaunp}).
\end{document}